\documentclass[12pt]{article}

\usepackage{epsfig}

\usepackage{amssymb}
\usepackage{amsfonts}

\def\be{\begin{equation}}
\def\ee{\end{equation}}
\def\ba{\begin{array}{c}}
\def\ea{\end{array}}

\newcommand{\kt}{\rangle}
\newcommand{\br}{\langle}

\begin{document}

%.
%
%\newpage

\begin{center}

{\Large

Generalized Bose-Hubbard Hamiltonians exhibiting a complete
non-Hermitian degeneracy

}

\vspace{0.8cm}

  {\bf Miloslav Znojil}

\vspace{0.2cm}

\vspace{1mm} Nuclear Physics Institute of the CAS, Hlavn\'{\i} 130,
250 68 \v{R}e\v{z}, Czech Republic

{e-mail: znojil@ujf.cas.cz}

\vspace{0.8cm}

\end{center}

\section*{Abstract}

The method of construction of the tridiagonal and symmetric
complex-matrix Hamiltonians $H^{(N)}(z)$ exhibiting an
exceptional-point (EP) degeneracy of the $N$th (i.e., maximal) order
at a preselected parameter $z=z^{(EPN)}=1$ is proposed and tested.
In general, the implementation of the method requires the use of
computer-assisted symbolic manipulations, especially at the larger
matrix dimensions $N$. The well known ${\cal PT}-$symmetric
$N$-by-$N$-matrix Bose-Hubbard Hamiltonians as well as their recent
$N\leq 5$ non-Bose-Hubbard alternatives as obtained as special
solutions. Several other $N\geq 6$ non-Bose-Hubbard models
$H^{(N)}(z)$ exhibiting the maximal EPN degeneracy are also
constructed and analyzed in detail. In particular, their
$z-$dependent, real or complex spectra of energies are displayed and
discussed near as well as far from $z^{(EPN)}$.

\subsection*{Keywords}

non-Hermitian quantum dynamics; multilevel degeneracies; exceptional
points of high orders;

\newpage

\section{Introduction}

In 2008, Graefe with coauthors \cite{Uwe} pointed out that one of
the most user-friendly features of the family \be
 H^{(2)}_{(BH)}(\gamma)=
 \left[ \begin {array}{cc} -i{\it \gamma}&1
 \\\noalign{\medskip}1&i{\it
 \gamma}
 \end {array} \right]\,,
 \ \ \ \
 H^{(3)}_{(BH)}(\gamma)=\left[ \begin {array}{ccc} -2\,i\gamma&
\sqrt{2}&0\\\noalign{\medskip}\sqrt{2}&0&
\sqrt{2}\\\noalign{\medskip}0&\sqrt{2}&2\,i\gamma\end {array}
\right]\,, \ldots
 \label{dopp2}
 \ee
of the special one-parametric ${\cal PT}-$symmetrized Bose-Hubbard
Hamiltonians (relevant, say, for the study of Bose-Einstein
condensation) is that their energy spectra are all known in closed
form,
 \be
 E_k^{(BH)}(\gamma)=(1-N+2k)\sqrt{1-\gamma^2}\,,
 \ \ \ \ k = 0, 1, \ldots, N-1\,.
 \label{espe}
 \ee
In spite of the manifest non-Hermiticity of the Hamiltonians, the
energy levels themselves remain real and non-degenerate (i.e.,
potentially observable \cite{ali}) inside an open and
dimension-independent interval of $\gamma \in (-1,1)$, therefore.

One of the related and phenomenologically as well as mathematically
most challenging problems emerges in the two limits of $\gamma \to
\pm 1$, i.e., on the boundary of the interval of stability and
potential observability of the energies. At these values, all of the
Hamiltonian matrices (\ref{dopp2}) cease to be diagonalizable,
acquiring the canonical form of the $N$ by $N$ Jordan block,
 \be
 H^{(N)}(\pm 1) \sim J^{(N)}(E)=\left [\begin {array}{ccccc}
    E&1&0&\ldots&0
 \\{}0&E&1&\ddots&\vdots
 \\{}0&0&E&\ddots&0
 \\{}\vdots&\ddots&\ddots&\ddots&1
 \\{}0&\ldots&0&0&E
 \end {array}\right ]\,
 \label{hisset}
 \ee
with $E=0$. This means that in the language of functional analysis
the two extreme values $\gamma=\pm 1$ of the parameter may be
interpreted as the Kato's exceptional points (EP, \cite{Kato}) of
order $N$ (EPN). In these two limiting cases the spectrum becomes
fully degenerate. In a broader context of general physics such a
specific EP extreme is often called non-Hermitian degeneracy
\cite{Berry}. In the narrower area of unitary quantum systems the
phenomenon is better known under the more intuitive, widely accepted
name of the spontaneous breakdown of ${\cal PT}$ symmetry
\cite{Carl}.

The authors of Ref.~\cite{Uwe} emphasized that one of the most
important phenomenological consequences of the existence of the EPN
$\gamma=\pm 1$ singularities should be seen in the variability of
the scenarios in which the degenerate spectrum ``unfolds'' under the
influence of perturbations. In {\it loc. cit.\,} the
phenomenologically most useful (viz., diagonal-matrix) choice of
these perturbations has even been shown tractable analytically, by
non-numerical means. A slightly more general version of the
perturbative model-building strategy has been discussed in
Ref.~\cite{admissible}.

The maximality of the non-Hermitian EPN degeneracy (\ref{hisset})
seems to follow from the highly specific choice of model
(\ref{dopp2}) and, in particular, from its Lie-algebraic origin and
symmetries \cite{Uwe}. In our recent paper \cite{45} we decided to
test such a conjecture. We managed to disprove it when, for the sake
of simplicity, we reparametrized $\gamma \to \sqrt {z} \in [0,1) $
(so that just a unique EPN value of $z^{(EPN)}=1$ had to be taken
into consideration), and when we only admitted the off-diagonal
deformations of the model. Even though we restricted our attention
just to the first two nontrivial, two-parametric deformations of the
respective original Bose-Hubbard Hamiltonians
$H^{(4,5)}_{(BH)}(\gamma)$ of Ref.~\cite{Uwe}, viz., to matrices
 \be
 H^{(4)}(z,A,B)= \left[ \begin {array}{cccc} -3\,i\sqrt {z}&\sqrt
 {B}&0&0
 \\\noalign{\medskip}\sqrt {B}&-i\sqrt {z}&\sqrt {A}&0
 \\\noalign{\medskip}0&\sqrt {A}&i\sqrt {z}&\sqrt {B}
 \\\noalign{\medskip}0&0&\sqrt {B}&3\,i\sqrt {z}\end {array}
 \right]\,
 \label{tripa}
 \ee
and
 \be
 H^{(5)}(z,A,B)=\left[
 \begin {array}{ccccc} -4\,i\sqrt {z}&\sqrt {B}&0&0&0
 \\\noalign{\medskip}\sqrt {B}&-2\,i\sqrt {z}&\sqrt {A}&0&0
\\\noalign{\medskip}0&\sqrt {A}&0&\sqrt {A}&0\\\noalign{\medskip}0&0&
\sqrt {A}&2\,i\sqrt {z}&\sqrt {B}\\\noalign{\medskip}0&0&0&\sqrt
{B}&4 \,i\sqrt {z}\end {array} \right]\,
 \label{petpa}
 \ee
we were able to conclude, due to the comparatively elementary nature
of such a generalization, that the maximal non-Hermitian EPN
degeneracy of the spectrum can be achieved,  at $z^{(EPN)}=1$, not
only via the conventional Bose-Hubbard choice of parameters
 \be
 A_{(BH)}^{(4)}=4\,,\ \ \ B_{(BH)}^{(4)}=3\,,\ \ \
 A_{(BH)}^{(5)}=6\,,\ \ \ B_{(BH)}^{(5)}=4\,
 \ee
but also via its symmetries violating and strongly deformed
alternatives
 \be
 A_{(non-BH)}^{(4)}=64\,,\ \ \ B_{(non-BH)}^{(4)}=-27\,,\ \ \
 A_{(non-BH)}^{(5)}=-54\,,\ \ \ B_{(non-BH)}^{(5)}=64\,.
 \label{twoex}
 \ee
Unfortunately, the applicability of the construction remained
restricted to the latter two examples. The method we used did not
seem to admit any immediate extension beyond $N=5$ (cf. section
\ref{Njegen} and paragraph \ref{Nje6} below). Still, the potential
physical relevance of the exceptional points of the higher orders
\cite{Muslimani} forced us to search for an amendment of the method.
The search succeeded, and its results will be reported in what
follows.

The overall idea of the innovated construction yielding the new,
non-BH Hamiltonians will be explained and, choosing $N=6$,
illustrated in section \ref{Njegen}. Several characteristic features
of its extension beyond $N=6$ will be then discussed in section
\ref{Njevic}. In subsequent section \ref{confluv} one of the key
technicalities (viz, the necessity of a reliable numerical proof of
the maximality of the degeneracy) will finally be identified and
resolved. The description of the parameter-dependence of the energy
spectra far from the EPN singularity will also be discussed, in
separate section \ref{hujus}, in some detail. After a thorough
discussion of some terminological and experimental aspects of our
results in section \ref{discussion}, the paper will be concluded by
a concise summary in section \ref{conclusions}.

\section{Three-parametric non-BH deformation at $N=6$
\label{Njegen}}

\subsection{The method of Ref.~\cite{45} and its
failure\label{Nje6}}

The feasibility of the constructions of the toy-model Hamiltonians
with property (\ref{hisset}) using the method of Ref.~\cite{45}
remained restricted to $N \leq 5$. Indeed, the next, $N=6$ model
with the three-parametric candidate
 \be
 H^{(6)}({ {z}},A,B,C)= \left[ \begin {array}{cccccc}
 -5\,i{{\sqrt {z}}}&\sqrt {C}&0&0&0&0
 \\
 \noalign{\medskip}\sqrt {C}&-3\,i{{\sqrt {z}}}&\sqrt {B}&0&0&0
 \\
 \noalign{\medskip}0&\sqrt {B}&-i{{\sqrt {z}}}&\sqrt {A}&0&0
 \\
 \noalign{\medskip}0&0
 &\sqrt {A}&i{{\sqrt {z}}}&\sqrt {B}&0
 \\\noalign{\medskip}0&0&0&\sqrt {B}&3\,i{{\sqrt {z}}}&
 \sqrt {C}\\\noalign{\medskip}0&0&0&0&\sqrt {C}&5\,i{{\sqrt {z}}}
 \end {array}
 \right]
 \label{sesti}
  \ee
for the deformed, non-Bose-Hubbard Hamiltonian may be assigned the
secular equation
%
%zg:=charpoly(ham6,sqrt(s));
%sufficient to study at overall scaling factor $z=1$:
 $$
 {s}^{3}+ \left( -A+35\,z-2\,C-2\,B \right) {s}^{2}+
 \left( {B}^{2}+2\,AC+
 28\,C\,z+259\,z^2+{C}^{2}-34\,A\,z+2\,BC-44\,B\,z \right) s+
 $$
 \be
 +30\,C\,z^2+{C}^{2}\,z-A{C}^{2}+
 25\,{B}^{2}\,z-225\,A\,z^2+150\,B\,z^2+225\,z^3+10\,BC\,z-30\,AC\,z=0\,
 \label{sestipo}
 \ee
where we abbreviated $E^2={s}$. Obviously, this equation only
defines all of the bound state energies $E_{\pm k}=\pm \sqrt{s_k}$,
$k=1,2,3$ in terms of the conventional but, in this particular
application, prohibitively complicated Cardano formulae for the
three roots $s=s_k$.

The localization of the EP6 non-Hermitian degeneracy becomes
difficult. The main problem is that the Cardano formulae define the
real roots as superpositions of complex numbers. This feature of the
construction (i.e., the necessity of a guarantee of the exact mutual
cancelation of the respective imaginary components of the root)
converts the construction of the canonical representation
(\ref{hisset}) of the Hamiltonians in question into a purely
numerical task. Unfortunately, such a numerical task is
ill-conditioned. In other words, the explicit
canonical-representation approach of Ref.~\cite{45} fails. Its
applicability remains restricted to the smallest matrix dimensions
$N \leq 5$. In what follows, an amended, ``implicit'', effective
alternative treatment of the problem will be developed and applied
at a few sample matrix dimensions $N>5$, therefore.

\subsection{Maximal degeneracy condition and the Gr\"{o}bner-basis
technique}

%{ }

%

The double symmetry (i.e., the symmetry with respect to its two main
diagonals) of our complex matrix (\ref{sesti}) is reflected by the
up-down symmetry $E_{\pm k}=\pm \sqrt{s_k}$ of the spectrum. Thus,
at any even matrix dimension $N=2J$, our key requirement of the
existence of the complete EPN degeneracy
 \be
 \lim_{z \to z^{(EPN)}}
 E_{\pm k}({ {z}},A^{(N)},B^{(N)},\ldots,Z^{(N)}) = 0\,,
 \ \ \ \ k = 1, 2, \ldots, J
 \label{degenery}
 \ee
(say, at the conveniently chosen $z^{(EPN)}=1$) implies that
the secular equation
  \be
 s^J + P_1(A,B,\ldots,Z)s^{J-1}+ \ldots +  P_{J-1}(A,B,\ldots,Z)s
 + P_J(A,B,\ldots,Z)=0
 \label{egen}
  \ee
must acquire, in the EPN limit, the utterly elementary form $s^J=0$.
In other words, the $J-$plet of the EPN-compatible parameters
$A^{(N)},B^{(N)},\ldots,Z^{(N)}$ must satisfy the $J-$plet of
polynomial equations
 \be
 P_{m}(A^{(N)},B^{(N)},\ldots,Z^{(N)})=0\,,\ \ \ \ m = 1, 2, \ldots, J\,.
 \label{greqs}
 \ee
The basic tool of the iterated-elimination solution of the similar
sets of equations is provided by the construction of the so called
Groebner basis. Such a construction is available via the
commercially available symbolic-manipulation software -- in what
follows we shall use MAPLE \cite{MAPLE}.

\subsection{EPN-admitting non-BH Hamiltonians at $N=6$}

Once we return to the $J=3$ secular equation (\ref{sestipo}) the
entirely routine application of the Gr\"{o}bner-basis solvers leads
to the five sets of solutions of the three coupled polynomial
equations (\ref{greqs}). The first, most elementary one just
reproduces the well known Bose-Hubbard matrix of Ref.~\cite{Uwe},
 $$
 (A^{(6)}_{(BH)},B^{(6)}_{(BH)},C^{(6)}_{(BH)}) = (9,8,5)\,.
 $$
In addition one reveals that there exist the other four sets of the
EP6-compatible solutions which are all expressed in terms of the
four roots $\xi$ of the following auxiliary polynomial equation
 \be
 416\,{\xi}^{4}+20909\,{\xi}^{3}+22505\,{\xi}^{2}
 +28734375\,\xi-48828125=0\,.
 \label{fourtho}
 \ee
For the sake of brevity we will drop here the discussion of the pair
of the complex roots. Thus, we will only consider the remaining two
real roots
 $$
 \xi_a=-65.80360706245132477179785808904814860530
 $$
and
 $$
 \xi_b=1.693394621288372898472626362413820064872\,.
 $$
In our calculations such a high-precision representation of these
``seed'' roots appeared necessary, mainly due to the perceivable
loss of numerical precision caused by the mutual cancelations
between the separate terms in the polynomials in question.

%
%%Pomoci toho vyjadrim cele reseni:
%
%
%                          Ba := -253.5822865
%%-253.5822865192720463817807750295375272977
%                           Aa := 673.7717872
%%673.7717871634467423071572662371713518059
%                          Ca := -65.80360706
%%-65.80360706245132477179785808904814860530
%
%> solve(fgb);

%The $J-$parametric $H^{(6)}(z,A,B,C)$
 %
% plot3d({1/s,zgb},z=5.488..5.4884,s=-108.4..-108.1,
%> view=-0.000001..0.000001,axes=framed,
%>   orientation=[269.9,0.4],numpoints=13000,tickmarks=[3,3,2]);
%plot3d({1/s,zga},z=-15.488..25.4884,s=-208.4..308.1,
%> view=-0.000001..0.000001,axes=framed,
%>   orientation=[269.9,0.4],numpoints=13000,tickmarks=[3,3,2]);

\begin{figure}[h]                    %instead of \begin{figure}[t]
\begin{center}                         %instead of \begin{center}
\epsfig{file=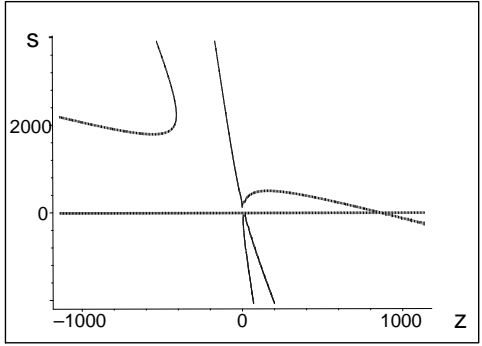,angle=270,width=0.36\textwidth}
\end{center}    % \sidecaption                      %instead of \end{center}
\vspace{2mm} \caption{The real roots $s=s(z)$ of secular
Eq.~(\ref{sestipo}) for Hamiltonian $H^{(6)}_{(a)}({ {z}})$.
 \label{gloja}
 }
\end{figure}

\subsubsection{The $a-$subscripted non-BH deformation\label{parag6}}

\begin{figure}[h]                    %instead of \begin{figure}[t]
\begin{center}                         %instead of \begin{center}
\epsfig{file=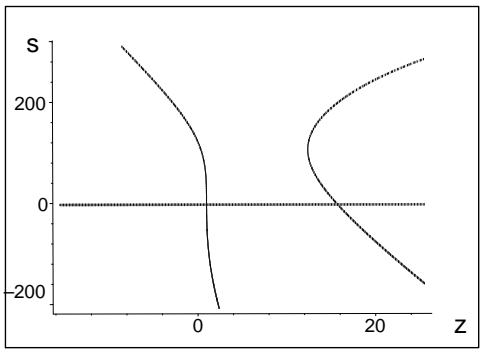,angle=270,width=0.36\textwidth}
\end{center}    % \sidecaption                      %instead of \end{center}
\vspace{2mm} \caption{The detail of avoided crossing in
Fig.~\ref{gloja} near $z^{(EP6)}=1$
 \label{magja}
 }
\end{figure}

In the first, $a-$subscripted case the conventional
Gr\"{o}bner-basis method gave us the desired EP6-compatible
parameters with numerical values
 \be
 A^{(6)}_a= 673.7717872\,,\ \ \
 B^{(6)}_a= -253.5822865\,,\ \ \
 C^{(6)}_a= -65.80360706\,.
 \label{asubs}
 \ee
This enables us to complement the $N=4$ and $N=5$ ``generalized
Bose-Hubbard'' models of Ref.~\cite{45} by their $a-$subscripted
$N=6$ descendant. The explicit form of its one-parametric
Hamiltonian $H^{(6)}_{(a)}(z)$ is obtained by the insertion of the
EP6-compatible parameters (\ref{asubs}) into the general
four-parametric matrix (\ref{sesti}). The global $z-$dependence of
its spectrum is displayed in Fig.~\ref{gloja} and, in a magnified
version near the EPN singularity $z^{(EP6)}=1$, in Fig.~\ref{magja}.

 %
%1. zleva: mensi dve jsou kladne ale mergujou uz v z cca -400, s cca 2000
%
%2. maximalni $s$ je realne zleva az po z=
%
%0.9999999999999999999999999999999999999993
%  %0.01041377580297735378670083806892428212825,
%
% a  pak jde dolu

%a druhy reseni je:
%
%%> cb:=fsolve(416*t^4-48828125+28734375*t+22505*t^2+20909*t^3,t,0..4);
%
%                          Cb := 1.693394621
%%1.693394621288372898472626362413820064872
%tj.
%                          Bb := -37.96027355
%%-37.96027355002016178540448838600535317215
%
%                           Ab := 107.5337579
%%107.5337578574635777738637240471830662145
%
%
%
%%
%
%a druhy reseni je:
%
%%> cb:=fsolve(416*t^4-48828125+28734375*t+22505*t^2+20909*t^3,t,0..4);
%
%                          Cb := 1.693394621
%%1.693394621288372898472626362413820064872
%tj.
%                          Bb := -37.96027355
%%-37.96027355002016178540448838600535317215
%
%                           Ab := 107.5337579
%%107.5337578574635777738637240471830662145

\subsubsection{The $b-$subscripted non-BH deformation}

In the second, $b-$subscripted case we arrive,
using the same recipe, at the other,
$b-$subscripted
set of the
EP6-compatible parameters
 \be
 A^{(6)}_b= 107.5337579\,,\ \ \
 B^{(6)}_b= -37.96027355\,,\ \ \
 C^{(6)}_b= 1.693394621\,.
 \ee
Along the same lines as before we also obtain the new one-parametric
Hamiltonian matrix $H^{(6)}_{(b)}(z)$ and the $z-$dependence of its
spectrum (see Figs.~\ref{bbegloja} and \ref{bbeg} and also the
comments in section~\ref{hujus} below).

%
%plot3d({1/s,zgb},z=-150.488..150.4884,s=-1080.4..1580.1,
%> view=-0.000001..0.000001,axes=framed,
%>   orientation=[269.9,0.4],numpoints=13000,tickmarks=[3,3,2]);
%
\begin{figure}[h]                    %instead of \begin{figure}[t]
\begin{center}                         %instead of \begin{center}
\epsfig{file=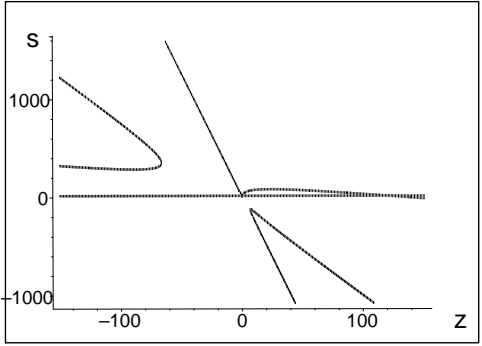,angle=270,width=0.36\textwidth}
\end{center}    % \sidecaption                      %instead of \end{center}
\vspace{2mm} \caption{Real roots $s(z)$ of secular
Eq.~(\ref{sestipo}) for Hamiltonian $H^{(6)}_{(b)}({z})$.
 \label{bbegloja}
 }
\end{figure}

Marginally let us add that for the fourth-order polynomial in
(\ref{fourtho}) the exact expression for the root can be obtained,
in closed form, via computer-assisted symbolic manipulations.
Unfortunately, this type of result is hardly of any use in practical
considerations. In contrast, the easy accessibility of the
approximate values of the two real roots $\xi_{a,b}$ can still help
us to study many relevant properties of the purely numerically
constructed EPN-admitting Hamiltonians.

\begin{figure}[h]                    %instead of \begin{figure}[t]
\begin{center}                         %instead of \begin{center}
\epsfig{file=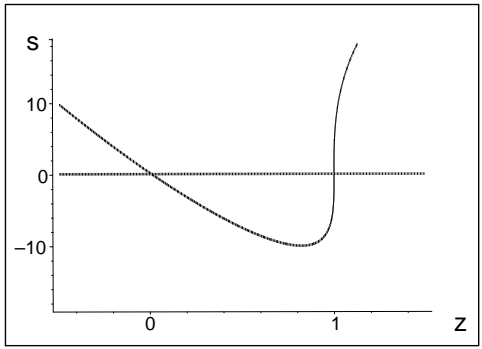,angle=270,width=0.36\textwidth}
\end{center}    % \sidecaption                      %instead of \end{center}
\vspace{2mm} \caption{The magnified shape of $s(z)$ near
$z^{(EP6)}=1$ in Fig.~\ref{bbegloja}.
 \label{bbeg}
 }
\end{figure}

%
%Once we return to the present choice of $N=6$ we may add that the
%numerical form of our pair of auxiliary roots $\xi_{a,b}$ could be
%replaced, in principle at least, by an entirely exact algebraic
%expression.

%\subsection{Hamiltonians $H^{(N)}$ with $N=7$ \label{Nje7}}

%\subsection{The $J-$parametric deformed Hamiltonians with odd $N$}

\section{General EPN-admitting models \label{Njevic}}

\subsection{The deformed-Hamiltonian ansatz at $N=7$}

The strategy of construction of the $J-$parametric deformed
Bose-Hubbard Hamiltonians with odd $N=2J+1$ remains unchanged. For
illustration let us consider the first nontrivial seven-dimensional
Hamiltonian ansatz
 $$
 H^{(7)}(z,A,B,C)=
 \left[ \begin {array}{ccccccc} -6\,i{\sqrt {z}}&\sqrt {C}&0&0&0&0&0
 \\
 \noalign{\medskip}\sqrt {C}&-4\,i{\sqrt {z}}&\sqrt {B}&0&0&0&0
 \\
 \noalign{\medskip}0&\sqrt {B}&-2\,i{\sqrt {z}}&\sqrt {A}&0&0&0
 \\
 \noalign{\medskip}0&0&\sqrt {A}&0&\sqrt {A}&0&0\\\noalign{\medskip}0
 &0&0&\sqrt {A}&2\,i{\sqrt {z}}&\sqrt {B}&0\\\noalign{\medskip}0&0&0&0&
 \sqrt {B}&4\,i{\sqrt {z}}&\sqrt {C}\\\noalign{\medskip}0&0&0&0&0&\sqrt {
 C}&6\,i{\sqrt {z}}\end {array} \right]\,.
 $$
We reveal that one of its eigenvalues is constant, identically equal
to zero, $E_0=0$. The rest of the spectrum, i.e., energies $E_{\pm
k}=\pm \sqrt{s_k}$ with $k=1,2,3$ are expressed again in terms of
the roots of the simplified secular equation
 $$
 {s}^{3}+ \left( -2\,B-2\,A-2\,C+56 \right) {s}^{2}+ \left( 2\,AB+{B}^{
2}+2\,BC+{C}^{2}+784-56\,B-104\,A+40\,C+4\,AC \right) s+
 $$
 \be
 +2304-2\,BAC+24
\,BC-96\,AC+72\,AB-2\,A{C}^{2}-1152\,A+576\,B+36\,{B}^{2}+4\,{C}^{2}+
192\,C=0\,.
\label{sedmipo}
 \ee
The mere inessential modification enables us now to apply the
above-outlined method of the search for the EPNs to the model with
$N=7$. Indeed, at any odd matrix dimension $N = 2J+1$, the condition
(\ref{degenery}) of the complete EPN degeneracy remains unchanged.
The same conventional choice of ${{z}}^{(EPN)}=1$ leads to the
formally identical constraint (\ref{greqs}). This specifies the
$J-$plet of the EPN-compatible parameters
$A^{(N)},B^{(N)},\ldots,Z^{(N)}$ at any $J$ and $N=2J+1$, in
principle at least.

\subsubsection{The $(\alpha,\beta)-$subscripted deformations}

Once we recall the Groebner-basis constructive technique we may
return to the present $N=7$ upgrade (\ref{sedmipo}) of the $J=3$
secular equation. This enables us to construct the EP7-supporting
Hamiltonians, indeed. Besides the known and expected Bose-Hubbard
solution with
 $$
 A^{(7)}_{(BH)} = 12\,,\ \ \  B^{(7)}_{(BH)} = 10\,,\ \ \
 C^{(7)}_{(BH)} = 6\,,
 $$
we discover the existence of the other two subfamilies of the
Hamiltonians.
%-0.69048105154699529125847122454416923479
In the first subfamily we arrive at the fixed integer value of
$A^{(7)} = -48$. The other two parameters become then defined, by
formulae
 $$
 B_{\alpha,\beta}^{(7)}=76-6\,r_{\alpha,\beta}\,,\ \ \ \
 C_{\alpha,\beta}^{(7)}=6\,r_{\alpha,\beta}\,,
 $$
in terms of the two respective non-numerical, exactly known roots
 \be
 r_{\alpha}=-59 + 10\,\sqrt{34}\approx -0.69048\,,
 \ \ \ \
  r_{\beta}=-59 - 10\,\sqrt{34}\approx -117.3095
  \label{elem}
 \ee
of quadratic equation $r^2+118\,r+81=0$.
%-117.3095189484530047087415287754558307652
%$ C = 6*RootOf(,label = _L2)$, $B = -6*RootOf(81+_Z^2+118*_Z,label = _L2)+76$,
%
% plus odmocnina: s real je minimalni, hyperboly jsou nad nim
%

\subsubsection{The $(\gamma,\delta)-$subscripted deformations}

In the second half of our exhaustive constructive analysis we would
have to analyze the second subfamily of the EP7-responsible coupling
constants. These appear to be defined in terms of the other pair of
the auxiliary roots
%3.791287847477920003294023596864004244492
%-0.791287847477920003294023596864004244492
 \be
 r_{\gamma}=3/2+1/2\,\sqrt {21}\approx 3.7912878\,,
 \ \ \ \
  r_{\delta}=3/2-1/2\,\sqrt {21}\approx -0.7912878
  \label{elemb}
 \ee
of the quadratic equation ${r}^{2}-3\,r-3=0$. The explicit closed
definitions of the respective EP7-compatible generalized, deformed
Bose-Hubbard Hamiltonians then read
 $$
 A_{\gamma,\delta}^{(7)}=36\,r_{\gamma,\delta}\,,
 \ \ \ \
 B_{\gamma,\delta}^{(7)}=28-54\,r_{\gamma,\delta}\,,
 \ \ \ \ C_{\gamma,\delta}^{(7)}=18\,r_{\gamma,\delta}\,.
 $$
The more detailed analysis of the
Hamiltonians defined by these parameters
is left to the readers.

%structure of the
%corresponding energy spectra
% $C = 18*RootOf(_Z^2-3*_Z-3,label = _L5)$, $B = -54*RootOf(_Z^2-3*_Z-3,label =
%_L5)+28$, $A = 36*RootOf(_Z^2-3*_Z-3,label = _L5)$

\subsection{The EPN constructions at $N>7$}

\subsubsection{The case of even $N=2J=8$ \label{Nje8}}

The occurrence of the EP8 degeneracy can be, naturally, studied
along the same lines as above. In the first step we have to evaluate
the secular polynomial and require that at ${ {z}}={ {z}}^{(EP8)}=1$
the secular equation degenerates again to its trivial EP8 form
$s^J=0$. This leads to the quadruplet of coupled polynomial
relations
 $$
 P_1=-A-2\,D+84-2\,C-2\,B=0\,,
 $$
 $$
 P_2=1974+2\,AC+2\,A D +{D}^{2}+50\,D+4\,B D
 +{B}^{2}+2\,C D +{C}^{2}-83\,A-142\,B+2\,BC-70\,C=0\,,
 $$
 $$
 P_3=1402\,C-2\,CB D +50\,{C}^{2}+12916-A{C}^{2}+74\,{B}^{2}
 +108\,BC+68\,AC+682\,D-2\,CA D -2\,{B}^{2} D-
 $$
 $$
 -152\,B D -A{D}^{2}+44\,C D -52
 \,A D -2\,B{D}^{2}+10\,{D}^{2}-2006\,B-1891\,A=0\,
 $$
and
 $$
 P_4=490\,BC+9\,{D}^{2}+630\,D-630\,A D +70\,{B}^{2}
 D +420\,B D +{B}^{2}{D}^{2}-9\,A{D}^{2}+
 $$
 $$
 +42\,C
 D +6\,B{D}^{2}+14\,CB D -42\,CA
 D  +11025+1225\,{B}^{2}+7350\,B-49\,A{C}^{2}-1470\,AC-11025\,A+
 $$
 $$
 +
 1470\,C+49\,{C}^{2}=0\,.
 $$
These equations are, expectedly, satisfied by the Bose-Hubard
parameters
 $$
 A^{(8)}_{(BH)} = 16, \ \  B^{(8)}_{(BH)} = 15,\ \
  C^{(8)}_{(BH)} = 12, \ \ D^{(8)}_{(BH)} = 7\,.
 $$
It is less elementary to find any other, non-Bose-Hubard solutions.
It was necessary to use the algebraic-manipulation software
\cite{MAPLE}. This enabled us to accept the computer-assisted
elimination strategy, to construct the Gr\"{o}bner basis and to
reduce the search for the EP8-supporting matrix elements of
$H^{(8)}$ to the purely numerical search for the roots of a single
polynomial $R(y)$ of degree $M(N)=M(8)=17$. The polynomial can still
be displayed in the single printed page, $R(y)=$

$=153712881941946532798614648361265167-
453762279414621179815552897029039797\,y+$

$+
235326754101824439936800228806905073\,{y}^{2}-
68875673245487669398850290405642067\,{y}^{3}+
$

$+
8129925258122948689157916436170874\,{y}^{4}-
145759836636885012145070948315366\,{y}^{5}+
$

$+
2361976444746440513605248930610\,{y}^{6}
+40525434802944282153115803370
\,{y}^{7}+
$

$+676326278232758784369966787\,{y}^{8}+
62429137451114251409236415\,{y}^{9}+
$

$+720991093724510065469933\,{y}^{10}
+14670346929744822064505\,{y}^{11}+
$

$+167556261648918275684\,{y}^{12}+
917318495163561932\,{y}^{13}+$

$+3133529909492864\,{y}^{14}+4574211144896
\,{y}^{15}-
$

$-5932158016\,{y}^{16}+314432\,{y}^{17}
$.

 \noindent
One should add that the search for the roots of this auxiliary
``resultant'' polynomial is a formidable numerical task. Using again
the restriction to the mere real roots we obtained the following
menu of seven eligible values
 $$
 -203.9147095411288,\ \  -156.6667001217788,\ \  -55.49992440658889,
 $$
 $$
 0.4192854385335118, \ \ 5.354156127796352, \ \ 1354.675194653849,
 \ \  18028.16789357534\,.
 $$
The insertion of any one of them would generate an independent
EP8-supporting Hamiltonian $H^{(8)}$. These insertions as well as
the derivation of their consequences remain routine.

%have already to be left to interested readers.

\subsubsection{The case of odd $N=2J=9$ \label{Nje9}}

At $N=9$ we have to deal with the quadruplet of polynomials

 $$
 P_1=-2\,C-2\,B+120-2\,D-2\,A\,,
 $$
 $$
 P_2=-88\,C+56\,D+4368+4\, {D} A-184\,B+{D}^{2}+{C}^{2}-232\,A +2\,
 {D} C+4\,B {D} +{B}^{2}+2\,BA+2\,CB+4\, CA\,,
  $$
 $$
 P_3=160\,CA-7808\,A+100\,{B}^{2}+68\,{C}^{2}-2\,A{C}^{2}+152\,CB-128\,
 {D} A-2\,A{D}^{2}-224\,B {D} +200\,BA+
 $$
 $$+1792
 \,D-2\,B{D}^{2}+52480+2752\,C+20\,{D}^{2}-3008\,B+72\,  D
  C-2\,CAB-4\, {D} AC-2\, {D} BC-4\,A
 B {D} -2\, {D} {B}^{2}\,,
  $$
 $$P_4=147456+6144\,D+12288\,C+2\,AB{D}^{2}-32\,A{D}^{2}+{B}^{2}{D}^{2}-73728
 \,A+36864\,B+1536\,B {D} +192\,AB {D}+
 $$
 $$
  +2304 \,{B}^{2}+2\,CAB D
   +256\,{C}^{2}+64\,{D}^{2}+16\,B{D}^{ 2}-3072\,  D A+96\, {D}
 {B}^{2}+
 $$
 $$
 +256\,
  {D} C-128\,A{C}^{2}+1536\,CB-6144\,CA+4608\,BA+32\,
  {D} BC-128\, {D} AC-128\,CAB\,.
  $$
% \noindent
The paradox of the simplification noticed after we moved from $N=2J$
to $N=2J+1$ at $J=3$ does not recur at $J=4$. The ``resultant''
polynomial $R(y)$ (again, of the 17th degree in $y$) does not
factorize in any obvious manner. The whole process as sampled at
$N=8$ must be repeated without any specific alterations. We omit the
details here.

%$N-$plets of

\section{The EPN confluence of {\em all\,} of the
eigenvectors\label{confluv}}

\subsection{The problem}

Up to now we only studied the necessary conditions of the complete
non-Hermitian degeneracy of the energy spectrum, $E_n \to
E^{(EPN)}=0$ at all $n=0,1,\ldots,N-1$. Such a result must be
complemented by the {\em elimination\,} of the {\em pathological\,}
possibility of the replacement of Eq.~(\ref{hisset}) by its
non-maximal-degeneracy alternatives like
 \be
  H^{(4)}(\pm 1) \sim J^{(2)}(E)\bigoplus J^{(2)}(E')
  =\left [\begin {array}{cccc}
  E&1&0&0
 \\
  0&  E&0&0
 \\{}0& 0&E'&1
 \\{}0& 0&0&E'
 \end {array}\right ]\,
 \label{43hisset}
 \ee
(with $N=4$ and $E=E'=0$)
or, in general, like
 \be
 H^{(N)}(\pm 1) \sim J^{(N_1)}(0)\bigoplus J^{(N_2)}(0)\,
 \label{pahisset}
 \ee
(with $N=N_1+N_2$
and
$ N_1\geq 1$ and $N_2\geq 1$),
or like
 \be
 H^{(N)}(\pm 1) \sim J^{(N_1)}(0)\bigoplus J^{(N_2)}(0)\bigoplus
 J^{(N_3)}(0)\,
 \label{3pahisset3}
 \ee
(with $N=N_1+N_2+N_3$
and
$ N_1\geq 1$, $N_2\geq 1$ and $N_3\geq 1$), etc.

In Ref.~\cite{45} the disproof of the existence of pathologies at
$z=z^{(EPN)}=1$ remained feasible, due to the absence of raounding
errors at $N=4$ and $N=5$, via the explicit construction of the
respective transition matrices $Q^{(N)}$. These matrices proved
obtainable directly from the definition
 \be
 H^{(N)}(z^{(EPN)}) Q^{(N)} = Q^{(N)} J^{(N)}(0)\,
 \label{2realt}
 \ee
of the canonical-representation mapping (\ref{hisset}). Such a
recipe necessarily fails in the presence of any rounding error,
i.e., as we saw, at any $N >5$. Even in the strictly EPN-compatible
scenario, every numerically evaluated matrix $H^{(N)}(z)$ with $z
\neq z^{(EPN)}$ then remains, in an arbitrarily small vicinity of
$z^{(EPN)}$, diagonalizable. In the language of functional analysis
one can say that the transition matrices $Q^{(N)}$ cease to exist at
almost all parameters. Any numerical attempt of their construction
must fail. The tests of the non-existence of the pathologies must be
indirect.

%
% \be
% H^{(N)}(A^{(EPN)}, \ldots, Z^{(EPN)}) Q^{(N)} = Q^{(N)} J^{(N)}\,.
% \label{realt}
% \ee
% Making simply the matrix $J^{(N)}$ in
%(\ref{realt}) block-diagonal, with a replacement, e.g., $J^{(N)}\to
%J^{(N_1)}\bigoplus J^{(N_2)}$ where $N=N_1+N_2$. Etc.

\subsection{The solution}

Due to the ubiquitous presence of the numerical uncertainties we
cannot construct the transition matrices $ Q^{(N)}$ which mediate
the isospectrality between our EPN-admitting matrix $H^{(N)}$ and
its Jordan-block partner $ J^{(N)}$. Formally this means that we
will not be able to solve Eq.~(\ref{2realt}). In the EPN regime of
interest, {\em any\,} standard numerical solver of such a problem
would become wildly unstable. As an immediate consequence we lose
the possibility of separating the EPN-related Hamiltonians from
their isospectral alternatives with the Jordan blocks in subspaces,
i.e., with a non-EPN, {\em incomplete\,} confluence of the
eigenvectors.

In place of the exact Hamiltonians  $H^{(N)}$ we only have access to
their numerical representations $H=H^{(N)}+V$ containing a random,
precision-dependent round-off perturbation $V={\cal O}(10^{-p})$. Up
to the set of measure zero this makes our perturbed EPN-compatible
Hamiltonians $H=H^{(N)}+V$, paradoxically, {\em diagonalizable}.
Naturally, all of their normalized eigenvectors $|\psi_n\kt$ are
mutually almost parallel. This is also the property which enables us
to test and confirm the occurrence of the EPN degeneracy even when
$V \neq 0$.

We imagined that in many implementations of computer arithmetics the
overall size of the round-off errors (i.e., of the exponent $p$ in
the estimate of $V={\cal O}(10^{-p})$) can be varied. Whenever we
amend the numerical precision $\sim {\cal O}(10^{-p})$ of the {\em
construction} of our EPN-supporting Hamiltonian $H^{(N)}(z)$, we
become able to distinguish between the correct, exhaustive, maximal
degeneracy scenario (cf. Eq.~(\ref{hisset})) and all of its
incorrect, pathological alternatives characterized by one of the
relations (\ref{43hisset}) or (\ref{pahisset}) or
(\ref{3pahisset3}), etc.

%\newpage

\begin{table}[h]
\caption{Numerical confirmation of the occurrence of EP6 in the
non-BH Hamiltonian $H^{(6)}_a(z)$. } \label{pexp4}
 \vspace{2mm}
  \centering
\begin{tabular}{||c||c|c||}
\hline \hline precision $p$ & min $\varrho_{mn}$ & max  $\varrho_{mn}$ \\
\hline
 \hline
  10& 7.2 $10^{-6}$ & 2.9 $10^{-5}$ \\
  20& 6.4 $10^{-7}$ & 2.6 $10^{-6}$ \\
  30& 2.7 $10^{-10}$ & 1.1 $10^{-9}$ \\
  40& 1.4 $10^{-13}$ & 5.5 $10^{-13}$ \\
\hline\hline
\end{tabular}
\end{table}

A sample of such a test is presented in Table~\ref{pexp4}.  In the
test the numerical matrix of the form $H=H^{(6)}_a+V$ (cf. paragraph
\ref{parag6}) with an unspecified, random round-off term $V={\cal
O}(10^{-p})$ was assigned the numerically evaluated normalized
eigenvectors $|\psi_n\kt$. Their mutual confluence was then measured
via an evaluation of the ``non-overlaps'' $\varrho_{mn}=1-|\br
\psi_m|\psi_n\kt|$. In the given example they are clearly decreasing
with the growth of $p$. The ``no pathology'' hypothesis
(\ref{hisset}) may be declared confirmed. The possibility of an {\em
incompleteness\,} of the degeneracy of the $N-$plet of the
eigenvectors of $H^{(N)}(z^{(EPN)})$ is persuasively excluded.

One should add that one of the important methodical merits of such a
$p-$variation strategy is that it is rather robust. As long as it
remains feasible at arbitrary matrix dimensions $N$, it enables us
to confirm the absence of the pathological alternative scenarios in
which the models with some hidden symmetry happen to be
block-diagonalizable.

\section{The spectra of energies $E=E(z)$ with $z \neq
z^{(EPN)}$\label{hujus}}

\subsection{$N=6$}

In both of the one-parametric generalizations (\ref{twoex})  of the
conventional one-parametric $N=6$ Bose-Hubbard model one encounters,
due to the absence of several conventional Lie-algebraic symmetries,
a richer structure of the $z-$dependence of the spectrum. Both of
the present deformed-model $N=6$ energy spectra can be perceived as
the phenomenologically welcome immediate complements of their $N<6$
predecessors of Ref.~\cite{45}.

\subsubsection{The $a-$subscripted Hamiltonian}

For the spectrum of the $a-$subscripted Hamiltonian one observes
that the whole real line of $z$ is split into several subintervals
of qualitatively different spectral form (cf. Fig.~\ref{gloja}). In
the leftmost subinterval, viz., for $z \in (-\infty,-z_1)$ with
$-z_1 \approx -400$, the values of $\gamma=\sqrt{z}$ are purely
imaginary. Hence,  it is not too surprising that all of the six
bound state energies remain strictly real.

At the emergent EP2 boundary $-z_1$ the innermost quadruplet of the
energies merges and complexifies, pairwise, at $s \approx 2000$.
Inside the subsequent interval $(-z_1,1)$, therefore, the innermost
quadruplet of the energies remains complex. We also find that the
remaining auxiliary root $s_3=E^2_{\pm 3}$ is positive.

The latter value becomes negative (and keeps decreasing) in the next
subinterval $(1, z_2)$ with $z_2 \approx 12.4$ (cf. the magnified
picture in Fig.~\ref{magja}). In this interval the related energies
$E_{\pm 3}$ become purely imaginary.

At the EP2-boundary $z_2$
one encounters again, at a positive value of $s \approx 104$, an
unfolding of the two real roots.
In the adjacent interval $(z_2,z_3)$ with $z_3 \approx 15.5$
the roots $s_1$ and $s_2$ stay real and positive. They also keep dominating
the third, negative real root $s_3$.
The
related
four energies are real.

At the end of the above interval the middle real root $s_2$ changes
sign. Inside the subsequent interval $(z_3,z_4)$ with $z_4 \approx
835$ the further two energies $E_{\pm 2}=\sqrt{s_2}$ acquire the
purely imaginary values. The maximal real root $s_1$ becomes also
negative beyond $z_4$. In the ultimate, rightmost interval of $z \in
(z_4,\infty)$ this converts the further two energies $E_{\pm
2}=\sqrt{s_2}$ into purely imaginary quantities as well.

 %
% 3. druhe dve remergujou pri z = cca 12.4 a s = cca 104
%
%% plot3d({zga},z=12.4..12.5,s=100..114,view=-0.000001..0.000001,axes=normal,
%%>   orientation=[269.9,0.4],numpoints=13000,tickmarks=[4,4,2]);
%
%
% 4. mensi ztraci kladnost pri z =
%
%
%       15.52763645951415873507489286950210212990
%
%5. vetsi taky klesne pod nulu v z=
%
%        835.0728226584407911295092711325673543554
%
%
%
%3. remergujou pri z = 5.488 a (zaporne) s =  -108.3 a jdou dolu

\subsubsection{The $b-$subscripted Hamiltonian}

Let us now turn attention to the other, $b-$subscripted Hamiltonian
and to its spectrum (cf. Fig.~\ref{bbegloja}). The observations and
conclusions remain similar. First of all, the leftmost EP2 boundary
gets merely shifted to $-z_1 \approx  =  -66.527$ yielding the
energy merger and subsequent complexification at $s \approx  346.2$.

In the second and third intervals $(z_1,z_2)$ and $(z_2,1)$ with
boundary $z_2 \approx 0.01041$ we notice the change of the shape of
the curve $s(z)$ near $z^{(EP6)}=1$. This is displayed in
Fig.~\ref{bbeg}.
%
%1. maximalni $s$ je realne: napred (zleva)  kladne, pak protina dolu v z=
%
%  0.01041377580297735378670083806892428212825,
%        131.6044071655022585572169986181292013718
The picture also implies the obvious classification of the energies.
Four of them  keep complex for $z \in (-z_1,z_3)$ with $z_3 \approx
131.6044$ while the remaining two remain real. In the subinterval $z
\in (z_2,1)$ they get again converted into the purely imaginary
quantities. Finally, in the rightmost interval of $z \in
(z_3,\infty)$ the model supports the two real and four purely
imaginary energy eigenvalues.
%
%
%2. zleva: druhe dve (jsou kladne a) mergujou v z =  -66.527, s =  346.2
%
%3. remergujou pri z = 5.488 a (zaporne) s =  -108.3 a jdou dolu

We may conclude that besides the preselected EP4 separator of the
real line of $z$, both of our ``a'' and ``b'' $N=6$ Hamiltonians
also appeared to keep the parallels with the $N=4$ results of
Ref.~\cite{45}. Thus, all of these generalized Bose-Hubbard-type
models are found to share the emergence of a few remote EP2
separators as well as of the qualitatively different types of
spectra supported by the individual, model-dependent subintervals of
$z$.

\subsection{$N=7$}

\begin{figure}[h]                    %instead of \begin{figure}[t]
\begin{center}                         %instead of \begin{center}
\epsfig{file=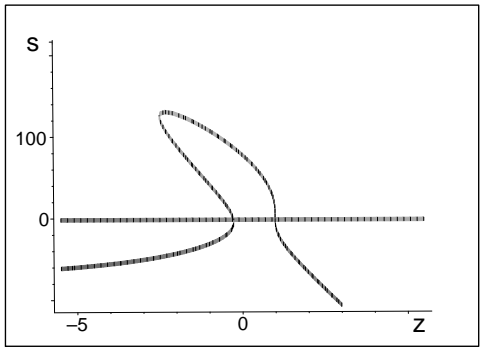,angle=270,width=0.36\textwidth}
\end{center}    % \sidecaption                      %instead of \end{center}
\vspace{2mm} \caption{The shape of real $s(z)$ near $z^{(EP7)}=1$ at
the first auxiliary root $r_\alpha$.
 \label{hheg}
 }
\end{figure}

 \noindent
In the vicinity of the most interesting EP parameter $z^{(EP7)}=1$
the local shape of the single real $\alpha-$subscripted curve
$s_\alpha (z)$ is shown in Fig. \ref{hheg}. In comparison with its
two $N=6$ predecessors (cf. Figs.~\ref{magja} and \ref{bbeg} above),
this shape is still different. The horizontal line which played just
the auxiliary, eye-guiding role at $N=6$ must be now reinterpreted
as representing also one of the energies.

\begin{figure}[h]                    %instead of \begin{figure}[t]
\begin{center}                         %instead of \begin{center}
\epsfig{file=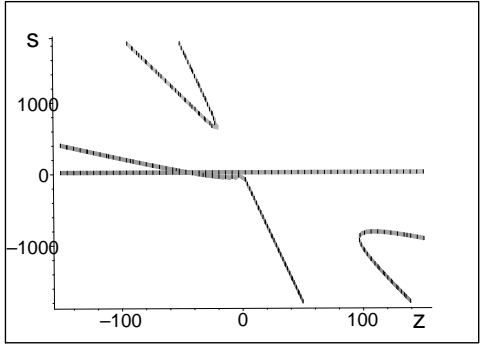,angle=270,width=0.36\textwidth}
\end{center}    % \sidecaption                      %instead of \end{center}
\vspace{2mm} \caption{The global shape of the real-root functions
$s(z)$ at the first auxiliary root $r_\alpha$ (cf. its small-$z$
detail in Fig.~\ref{hheg}).
 \label{bhheg}
 }
\end{figure}

In a more detailed analysis of the spectrum the points of the change
of the sign of $s_\alpha(z)$ can be also localized exactly. In
Fig.~\ref{hheg} we have $s_\alpha(z)=0$ not only at the EPN
prescribed value $z=1$ but also at the small negative exact nodal
point
$$z_+=5\,\sqrt {34}-{\frac {103}{2}}+\frac{5}{2}\,
\sqrt {381-52\,\sqrt {34}}\approx -0.2955.$$
%095937354074225086194406509450862971
In the global picture of Fig.~\ref{bhheg} we also find the third
real zero of $s_\alpha(z)$ at $z_-\approx -44.39497$.
%145781158786874985178389322414851

% dd je globalni pohled
%
%plot3d({1000/s,zga},z=-150.488..150.4884,s=-1808.4..1808.1,
%>  view=-0.01..0.01,axes=framed,
%>   orientation=[269.7,0.8],numpoints=3000,tickmarks=[3,3,2]);
% ddd je globalni pohled
%
%% plot3d({zga},z=-50.488..150.4884,s=-2008.4..2008.1,
%> view=-0.01..0.01,axes=framed,
%>   orientation=[269.7,0.8],numpoints=3000,tickmarks=[3,3,2]);

%hhh je to prave okoli

%plot3d({1/s,zga},z=-5.488..5.4884,s=-108.4..208.1,
%> view=-0.001..0.001,axes=framed,
%>   orientation=[269.7,0.8],numpoints=3000,tickmarks=[3,3,2]);

%lokalne wwww, globalne ww

\begin{figure}[h]                    %instead of \begin{figure}[t]
\begin{center}                         %instead of \begin{center}
\epsfig{file=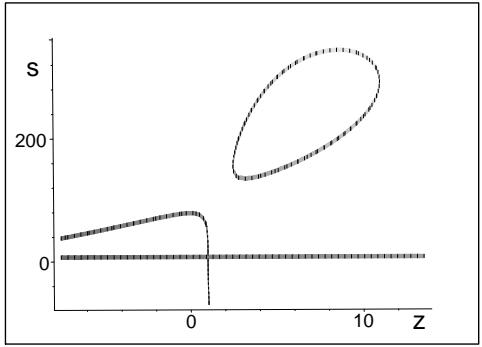,angle=270,width=0.36\textwidth}
\end{center}    % \sidecaption                      %instead of \end{center}
\vspace{2mm} \caption{The shape of $s(z)$ near $z^{(EP7)}=1$ at the
auxiliary root $r_\beta$.
 \label{mbbeg}
 }
\end{figure}

%s minjusem ovsem jinak
Naturally, after the change $\alpha \to \beta$ of sign in
Eq.~(\ref{elem}) the zeros of the real function $s_\beta(z)$ get
shifted,
$$
z_+ \approx -0.69048\,,\ \ \ \  z_-= \approx -146.048\,.
$$
%-0.69048105154699529125847122454416923479
%-146.0482558314692664161951024738123759485
A surprise still emerges at the second auxiliary root $r_\beta$.
Qualitatively, the global shape of $s_\beta(z)$ parallels
Fig.~\ref{bhheg} -- only a rescaling of the axes makes the main
difference. In contrast, locally, a magnified detail as given in
Fig.~\ref{mbbeg} shows a different, utterly unexpected new pattern
near $z^{(EP7)}=1$ when compared with Fig.~\ref{hheg}.

\section{Discussion\label{discussion}}

%The tendency was reflected, in
%particular, by our present study of a specific class of
%non-Hermitian Hamiltonians. These constructive results may be now
%complemented by a few additional physics-oriented comments.

\subsection{Complex symmetric Hamiltonians}

In the Heisenberg's more than ninety years old formulation of
quantum mechanics \cite{Heisenberg} an important role has been
played by the finite-dimensional and real symmetric matrices
$H=H^T$. Nowadays, quantum physics still relies heavily upon the
inspiration provided by such an elementary mathematics. In the
Kato's influential monograph \cite{Kato}, for example, several
abstract features of various advanced quantum Hamiltonians may still
be found illustrated by the most elementary two by two matrices,
real or complex, and symmetric, or not. Typically, multiple relevant
physical questions (e.g., of the stability of quantum systems under
small perturbations) as well as some related mathematical concepts
may be found explained there in terms of these utterly elementary
toy models.

The recent return of interest to less conventional elementary models
(cf., e.g., the use of  {\em complex symmetric\,} matrices in
Refs.~\cite{dva,eva,ctyri}) found its motivation in an innovative
physical interpretation of EPs. Specifically, we may mention the
relevance of EPs in the study of classical as well as quantum
dynamical systems~\cite{Milburn,Doppler,catast} and/or the key role
of the unfolding of the EPs of the $N-$th order in the description
of the dynamics of the Bose-Einstein condensation \cite{Uwe}.

What is observed in the new research motivated by similar ideas is
not only an enormous increase in the experimental activities (cf.,
e.g., their concise list in \cite{review}) but also the growth of
quality of theoretical investigations \cite{book}. Typically, the
interest of mathematicians in the properties of finite-dimensional
complex symmetric matrices \cite{Craven} is now being extended to
the infinite-dimensional cases \cite{Garcia} and to the various
non-Hermitian operators exhibiting the innovative Krein-space
self-adjointness {\it alias\,} ${\cal PT}-$symmetry
\cite{ali,Carl,book}. In parallel, the accessibility of the
efficient, computer-assisted symbolic manipulation techniques opens
also new horizons in our understanding of certain less conventional
finite-dimensional models (cf., e.g., the lists of references in
\cite{Uwe,Gegenbauer,Geyer}).

\subsection{Bose-Hubbard and non-Bose-Hubbard models}

The latter studies inspired our theoretical investigation \cite{45}
where we considered the tridiagonal, ${\cal PT }-$symmetric and
complex-symmetric $N$ by $N$ matrix Hamiltonians
$H^{(N)}(A,B,\ldots,Z)$ up to $N=5$. The availability of the utterly
elementary formulae for the multiparametric energies
$E_n^{(N)}(A,B,\ldots,Z)$ enabled us to succeed in obtaining an
exhaustive $N \leq 5$ classification of the subset of models
$H^{(N)}(A,B,\ldots,Z)$ guaranteeing, at an {\it ad hoc} set of
parameters $A^{(EPN)},B^{(EPN)},\ldots,Z^{(EPN)}$, the existence and
guarantee of the confluence of {\em all\,} of the lower-order EPs
into a single, maximal EPN singularity.

In our present paper we succeeded in extending these results to all
$N$. The level of the complexity of our innovated, unusual
Hamiltonians and of their spectra appears controlled, roughly
speaking, by the integer part $J$ of $N/2$. A remark should be added
concerning the presentation of the results at the larger dimensions
$N$ when all formulae become rather long, not fitting a printed-page
format. We, nevertheless, imagined that a detailed discussion of the
models with $N\leq 9$ already provides a sufficiently comprehensible
picture of the situation and, in particular, of the universality of
our construction of the generalized Bose-Hubbard-type Hamiltonians,
exhibiting still the presence of the $N-$th order exceptional-point
degeneracy.

Among the current applications of the similar models in physics one
can notice an intensification of interest in the so called open
quantum systems in which one admits a non-unitarity of the evolution
caused by an uncontrolled interaction with an environment
\cite{Nimrod}. It is not too surprising that one of the most natural
theoretical formulations of such a situation is offered by a
combination of the EP-related phenomenology with the real- or
complex-symmetric-matrix mathematics. {\it Pars pro toto\,} we
should mention here a well-balanced combination of the mathematical
and phenomenological insight provided by the Bose-Hubbard
multi-bosonic models studied in both of their real- and
complex-symmetric versions -- cf., e.g., \cite{Uweb} and
\cite{Uwec}, respectively.

\subsection{Non-equivalent concepts of non-Hermitian physics}

%in the
%light of the Stone's theorem
%
%\subsection{Non-Hermitian Bose-Hubbard Hamiltonians
%and their generalizations\label{app}}

In conventional quantum mechanics the predictions  concerning an
observable $\Lambda$ are probabilistic, expressed in terms of matrix
elements $\br \psi(t)|\Lambda|\psi(t)\kt$. The time-dependence of
the states $|\psi(t)\kt$ is controlled by Schr\"{o}dinger equation
 \be
 {\rm i}\partial_t\,|\psi(t)\kt = H\,|\psi(t)\kt
 \,.
 \label{seqa}
 \ee
Traditionally, the evolution is assumed unitary so that the
Hamiltonian itself must be chosen selfadjoint, $H = H^\dagger$
\cite{Stone}. A less conventional, non-unitary-evolution paradigm in
which $H \neq H^\dagger$ is being developed in several alternative
directions at present \cite{book}. Opening several entirely new
areas of research. Thus, whenever one speaks about non-Hermitian
Hamiltonians (i.e., about operators $H$ such that $H \neq H^\dagger$
in some preselected Hilbert space ${\cal L}$), it is necessary to
check their spectrum $\sigma(H)$. This enables us to distinguish
between the so called quasi-Hermiticity (best known and used in
nuclear physics \cite{Geyer}, with $\sigma(H) \in \mathbb{R}$) and
the less unusual genuine non-Hermiticity taking place whenever
$\sigma(H)$ is {\em not\,} real \cite{Nimrod}.

In our present paper our attention was concentrated on the
open-system scenario in which the Hilbert space ${\cal L}$ of ket
vectors $|\psi(t)\kt$ is considered physical. For this reason the
non-unitarity of the evolution (caused by the non-Hermiticity of
$H$) is accepted as natural, finding its explanation in a
non-negligible interaction of the quantum system in question with a
certain rather vaguely specified environment (cf., e.g., the
original ideas of Feshbach \cite{Feshbach} and L\"{o}wdin
\cite{Feshbachb} as well as their multiple most recent theoretical
developments and implementations as sampled, say, in reviews
\cite{Nimrod,Ingrid}).

\subsection{Exceptional points: Theory vs. experiment}

The Kato's \cite{Kato} concept of exceptional points found multiple
applications in the study of Hamiltonians which are non-self-adjoint
in a preselected Hilbert space ${\cal L}$, physical or not.
Typically, the pioneering simulation of the confluence of two
quantum resonances in \cite{dva} offered one of the first {\em
experimental\,} localizations of the EP2 anomaly in a classical
microwave setup. What followed was a long series of other
experimental studies of EP2s covering, say, their manifestations in
electronic circuits \cite{xx22}, in exciton-polariton billiards
\cite{xx24} or in optomechanical systems \cite{xx26}. Not too
surprisingly, a ``natural'' transition to the analogous experimental
EPN scenarios with $N \geq 3$ opened a number of new and challenging
questions and obstacles \cite{Cart2}. In fact, one of the first
physical systems making the EP3-related patterns experimentally
accessible was only proposed, in the pioneering study \cite{eva}, in
2012.

Also the progress in the underlying theory of the localization of
EPNs is not too quick. Its characteristic nontrivial $N=3$
implementation as proposed in \cite{Cart} consisted, for example, of
a mere triplet of coupled wave-guides, with the gain and loss areas
arranged in a ${\cal PT }-$symmetric manner. An experimentally
feasible arrangement of the semiconductor wave-guides has been
determined. The most elementary tridiagonal and complex symmetric
choice of the matrix $H^{(3)}$ assigned to the system appeared
realistic and sufficient for the purpose.

The idea of the study of the role of EPs in Bose-Hubbard-type models
was followed also in \cite{45}. The present further development and
completion of such a project was still motivated by the generic
phenomenological appeal of the models with higher-order EPs as well
as by the feasibility of an experimental realization and
verification of the EP-related theoretical hypotheses and measurable
predictions. A short outline of the history was provided here in
Introduction. In our present paper we only re-emphasized the
relevance of the higher-order EPs and, in particular, of the
guarantee of their occurrence in the fairly broad class of the
generalized, Bose-Hubbard-inspired $N$ by $N$ matrix Hamiltonians
$H^{(N)}$.

\section{Summary\label{conclusions}}

Our present extension of the non-BH constructions beyond $N= 5$ was
achieved only after a thorough innovation of the method. The main
obstacle appeared represented by the presence of the round-off
errors in the numerically constructed Hamiltonian matrices
$H_{(non-BH)}^{(N)}(z)$ at the larger dimensions. Originally we were
sceptical in this respect. A decisive progress has only been
achieved when we turned attention to the computer-assisted
constructions using the variable-precision arithmetics. Without an
explicit construction of transition matrices $Q^{(N)}$, this enabled
us to check the existence and maximality of the EPN anomaly.

In its present form, our arbitrary$-N$ construction of the non-BH
Hamiltonians still originates from the energy-determining secular
equation. The energy-degeneracy constraint (\ref{degenery}) then
yields the polynomial constraints equivalent to a nonlinear
algebraic set of $J$ coupled polynomial equations for $J$ unknown
quantities $A^{(EPN)},B^{(EPN)}$, $\ldots$, $Z^{(EPN)}$, real or
complex [cf. Eq.~(\ref{greqs})]. Such a set is, in general, solvable
by the well known Gr\"{o}bner-basis elimination technique. Thus, at
any Hamiltonian-matrix dimension $N$ the Gr\"{o}bner-basis
elimination reduces the construction of the parameters
$A^{(EPN)},\ldots$ to the localization of the roots $y_k$ of a
single polynomial $R^{(N)}(y)$.

The viability of such an approach and algorithm has been confirmed
here to work very comfortably up to the fairly large matrix
dimensions $N$. A decisive technicality has been found to lie in the
localization of the ``seed'' roots $y_k$. The reason is that the
degree $M(N)$ of polynomial $R^{(N)}(y)$ grows quickly with $N$. For
example, we had $M(N)=17$ in our $J=4$ samples of dimensions $N=8$
and $N=9$. Thus, the main criterion of the practical feasibility of
the construction is given by the value of $M(N)$.

Incidentally, the quick increase of $M(N)$ makes it practically
impossible to display any sample results in print at $N \geq 10$.
Even the very length of the individual (integer) numerical
coefficients in $R^{(N)}(y)$ would exceed the single-line capacity
in such a case. Fortunately, whenever needed, it makes sense to keep
the $N \geq 10$ results stored in the computer while asking just for
the display of output containing the measurable (e.g., spectral)
predictions.

In the physical, phenomenological model-building context the most
characteristic appeal of our present EPN-supporting toy-model
Hamiltonians is twofold. Firstly, these models offer a truly rich
menu of the parameter-dependence of the energy spectra even far from
the EPN merger. Secondly, the price to be paid for the flexibility
remains acceptable. Thus, the key message delivered by our study is
that one can deal with the non-BH Hamiltonian matrices at the
unexpectedly large dimensions $N$. The way to the success has been
found in keeping the influence of the round-off errors under
control. This was shown to confirm that in the EPN limit of $z \to
1$ the non-BH quantum systems really {\em do\,} reach the {\em
maximal\,} non-Hermitian degeneracy at which {\em all\,} of the
eigenvectors of the Hamiltonian {\em do\,} simultaneously merge.

\section*{Acknowledgement}

Work supported by GA\v{C}R Grant Nr. 16-22945S.

%\end{document}

%i.e., with $N=2J$ or $N=2J+1$ where $J=3$


\newpage

\begin{thebibliography}{99}
%\footnotesize\itemsep=0pt

\bibitem{Uwe}
E. M. Graefe, U. G\"{u}nther, H. J. Korsch and A. E. Niederle,
%. A non-Hermitian PT symmetric Bose-
%Hubbard model: eigenvalue rings from unfolding higherorder
%exceptional points. J Phys A 41(25):255206, 2008.
%doi:10.1088/1751-8113/41/25/255206.
J. Phys. A: Math. Theor 41, 255206 (2008).



\bibitem{ali}
A. Mostafazadeh,
%Pseudo-Hermitian Quantum Mechanics,
%%
%arXiv:0810.5643,
%.
% arXiv:0810.5643 [pdf, ps, other]
%Ali Mostafazadeh Comments: 76 pages, 2 figures, 243 references,
%revised version
Int. J. Geom. Meth. Mod. Phys. 7, 1191
%-1306
 (2010).

\bibitem{Kato}
T. Kato, Perturbation Theory for Linear Operators (Springer-Verlag,
Berlin, 1966).

\bibitem{Berry}
M. V. Berry, Czech. J. Phys. 54, 1039
%-1048
(2004).

\bibitem{Carl}
%C. M. Bender, D. C. Brody and H. F. Jones,
%Phys. Rev. Lett. 89, 270401 (2002) and Phys. Rev. D 70, 025001 (2004);
%
C. M. Bender,
%"Making Sense of Non-Hermitian Hamiltonians"
Rep. Prog. Phys. 70, 947
%-1018
(2007);
%-1018."Making Sense of Non-Hermitian Hamiltonians"
%C. M. Bender Invited review article published in Reports on Progress
%in Physics 70, 947-1018 (2007) [arXiv: hep-th/0703096]

C. M. Bender,
PT Symmetry
In Quantum and Classical Physics (World Scientific, Singapore, 2019);
edited book with
%https://doi.org/10.1142/q0178 | January 2019
%Pages: 468
%By (author): C. M. Bender (Washington University in St. Louis, USA)
%C
contributions by P. E. Dorey, C. Dunning, A. Fring, D. W. Hook,
H. F. Jones, S. Kuzhel, G. L\'{e}vai and R. Tateo.

\bibitem{admissible}
A. P. Seyranian and  A. A. Mailybaev, Multiparameter Stability
Theory with Mechanical Applications (World Scientific, Singapore,
2003);

M. Znojil,
%Admissible perturbations and false instabilities in
%PT-symmetric quantum systems. .
Phys. Rev. A 97, 032114 (2018).
% (13
%pages, No 3, 1 March) DOI: 10.1103/PhysRevA.97.032114
%(arXiv:1803.01949).

\bibitem{45}
M. Znojil,
% Complex symmetric Hamiltonians and exceptional points of order four and five
 Phys. Rev. A 98, 032109 (2018).
% DOI: 10.1103/PhysRevA.98.032109 , jour. No. 3, published 13 September
% (arXiv:1808.07472).
%%
%\bibitem{45}
%M. Znojil, Phys. Rev. A 98, 032109 (2018).

\bibitem{Muslimani}
%LETTER Reverse PT phase transition across exceptional points of any
%order Mohammad H. Teimourpour1, Sahin K. Özdemir2 and Ramy
%El-Ganainy1 Published 23 October 2017 • Copyright © EPLA, 2017 EPL
%(Europhysics Letters), Volume 119, Number 3
M. H. Teimourpour1, S. K. \"{O}zdemir and R. El-Ganainy,
EPL 119, 34003 (2017);

%
%
%Optical lattices with higher-order exceptional points by
%non-Hermitian coupling By:Zhou, XP (Zhou, Xingping)[ 1,2 ] ; Gupta,
%SK (Gupta, Samit Kumar)[ 1,3 ] ; Huang, Z (Huang, Zhong)[ 1,2 ] ;
%Yan, ZD (Yan, Zhendong)[ 1,2 ] ; Zhan, P (Zhan, Peng)[ 1,2,4 ] ;
%Chen, Z (Chen, Zhuo)[ 1,2,4 ] ; Lu, MH (Lu, Minghui)[ 1,3,4 ] ;
%Wang, ZL (Wang, Zhenlin)[ 1,2,4 ] APPLIED PHYSICS LETTERS Volume:
%113  Issue: 10 Article Number: 101108 DOI: 10.1063/1.5043279
%Published:SEP 3 2018 Document Type:Article View Journal Impact
%Abstract Exceptional points (EPs) are degeneracies in open wave
%systems with coalescence of at least two energy levels and their
%corresponding eigenstates. In higher dimensions, more complex EP
%physics not found in two-state systems is observed. We consider the
%emergence and interaction of multiple EPs in a four coupled optical
%waveguides system by non-Hermitian coupling showing a unique EP
%formation pattern in a phase diagram. In addition, absolute phase
%rigidities are computed to show the mixing of the different states
%in definite parameter regimes. Our results could be potentially
%important for developing further understanding of EP physics in
%higher dimensions via generalized paradigm of non-Hermitian coupling
%for a generation of parity-time devices. Published by AIP
%Publishing.
X.-P. Zhou, S. K. Gupta,
Z. Huang, Z. D. Yan, P. Zhan, Z. Chen, M.-H. Lu
and Z.-L. Wang,
Appl. Phys. Lett. 113, 101108 (2018).

\bibitem{MAPLE}
B. W. Char et al, Maple V (Springer, New York, 1991).


\bibitem{Heisenberg}
W. Heisenberg, Z. Phys. 33, 858-888 (1925).



\bibitem{dva}
M. Philipp, P. von Brentano, G. Pascovici, A. Richter,
%Frequency and width crossing of two interacting resonances in a
%microwave cavity.
Phys. Rev. E 62, 1922-1926 (2000);
%doi:10.1103/PhysRevE.62.1922.
%%

%Exceptional points enhance sensing in an optical microcavity Weijian
W.-J. Chen, S. K. \"{O}zdemir, G.-M. Zhao, J. Wiersig and L. Yang,
Nature  548, 192
%- 196
(2017).


\bibitem{eva}
G. Demange and E.-M. Graefe,
%. Signatures of three coalescing
%eigenfunctions. J Phys A 45(2):025303, 2012.
%doi:10.1088/1751-8113/45/2/025303. To cite this article: Gilles
%Demange and Eva-Maria Graefe 2012 J. Phys. A: Math. Theor. 45 025303
%
%
%Demange, G., Graefe, E.-M.: Signatures of three coalescing
%eigenfunctions.
J. Phys. A: Math. Theor 45, 025303 (2012).

\bibitem{ctyri}
M. Znojil,
%205. {236, [189]} Miloslav Znojil, Determination of the domain of
%the admissible matrix elements in the four-dimensional PT-symmetric
%anharmonic model.
Phys. Lett. A 367, 300 (2007);
%-306
%(quant-ph/0703168).

%Atushi
A. Tanaka,
%Sang Wook
S. W. Kim and T. Cheon.
 Phys. Rev. E 89,
042904
%– Published 7 April 2014
% PHYSICAL REVIEW E 89, 042904
(2014);

%Kun
K. Ding,
%Guancong
G.-C. Ma,
%Meng
M. Xiao, Z.-Q. Zhang and C.-T. Chan,
Phys. Rev. X 6, 021007 (2016).
%
%Emergence, Coalescence, and Topological Properties of Multiple
%Exceptional Points and Their Experimental Realization
%
% Phys.
%Rev. X 6, 021007 – Published 12 April 2016


%\bibitem{Heiss}
%A. P. Seyranian and  A. A. Mailybaev,
%Multiparameter Stability
%Theory with Mechanical Applications
%(World Scientific, Singapore, 2003);
%
%M. V. Berry,
%% Physics of non-Hermitian degeneracies.
%Czech. J. Phys. 54, 1039
%%- 1047
%(2004);
%
%W. D. Heiss, Czech. J. Phys. 54, 1091 (2004);
%
%W. D. Heiss, J. Phys. A: Math. Theor. 45, 444016 (2012);
%
%V. M.
%Martinez Alvarez, J. E. Barrios Vargas and L. E. F. Foa Torres,
%Phys. Rev. B 97, 121401(R)
%%– Published 7 March
%(2018).
%%

\bibitem{Milburn}
T. J. Milburn, J. Doppler, C. A. Holmes,
S. Portolan, S. Rotter and
P. Rabl,
%General description of quasi-adiabatic dynamical phenomena near
%exceptional points
%
%Thomas J. Milburn, Jörg Doppler, Catherine A. Holmes, Stefano
%Portolan, Stefan Rotter, Peter Rabl
    Phys. Rev. A 92, 052124 (2015).
%DOI:    10.1103/PhysRevA.92.052124 Cite as:    arXiv:1410.1882
%[quant-ph]

\bibitem{Doppler}
J. Doppler, A. A. Mailybaev, J. B\"{o}hm, U.
Kuhl, A.
Girschik, F. Libisch, T. J. Milburn, P. Rabl, N. Moiseyev and S.
Rotter,
% arXiv:1603.02325 [pdf]
%Dynamically encircling exceptional points in a waveguide: asymmetric
%mode switching from the breakdown of adiabaticity
%
%Jörg Doppler, Alexei A. Mailybaev, Julian Böhm, Ulrich Kuhl, Adrian
%Girschik, Florian Libisch, Thomas J. Milburn, Peter Rabl, Nimrod
%Moiseyev, Stefan Rotter Journal-ref:
Nature 537, 76 (2016).
%-79



\bibitem{catast}
M. Znojil,
% M. Maximal couplings in PT-symmetric chain-models with the
% real spectrum of energies.
J. Phys. A: Math. Theor. 40, 4863
%- 4875
(2007);

M. Znojil,
%"Quantum catastrophes: a case study."
J. Phys. A: Math. Theor. 45, 444036  (2012);
%http://stacks.iop.org/1751-8121/45/444036
%doi:10.1088/1751-8113/45/44/444036 arXiv: 1206.6000

G. L\'{e}vai, F. R\r{u}\v{z}i\v{c}ka and M. Znojil
%Three solvable matrix models of a quantum catastrophe
Int. J. Theor. Phys. 53, 2875
% - 2890); arXiv: 1403.0723
(2014);


M. Znojil,
%"Solvable model of quantum phase transitions and the
%symbolic-manipulation-based study of its multiply
%degenerate exceptional points and of their unfolding."
Ann. Phys. (NY) 336, 98  (2013);
%-111
%http://dx.doi.org/10.1016/j.aop.2013.05.016
%(arXiv:1305.4822 [quant-ph])


M. Znojil,
%: "Parity-time symmetry and the toy models of gain-loss
%dynamics near the real Kato's exceptional points."
Symmetry 8, 52 (2016).
%52 (doi: 10.3390/sym8060052) (open-access Special Issue
%``Parity-Time Symmetry in Optics and Photonics" )





\bibitem{review}
%Non-Hermitian physics and PT symmetry By:
R. El-Ganainy, K. G. Makris, M. Khajavikhan et al., Nat. Phys. 14,
11 (2018).



\bibitem{book}
%F. Bagarello et al, Eds,
%``Non-Selfadjoint Operators in Quantum Physics:
%Mathematical Aspects'',  Wiley, Hoboken,
%2015.
%
F. Bagarello, J.-P. Gazeau, F. H. Szafraniec and M. Znojil, Eds.,
Non-Selfadjoint Operators in Quantum Physics: Mathematical Aspects
(Wiley, Hoboken, 2015).



\bibitem{Craven}
%Complex symmetric matrices
B. D. Craven, J. Austral Math. Soc. 10, 341 (1969).
%Journal of the Australian Mathematical Society
% Volume 10, Issue 3-4 November 1969 , pp. 341-354


\bibitem{Garcia}
%STEPHAN RAMON GARCIA AND MIHAI PUTINAR
S. R. Garcia and M. Putinar,
Trans. Amer. Math. Soc. 358, 1285 (2005).


\bibitem{Gegenbauer}
M. Znojil, Phys. Rev. A 82, 052113 (2010);
%Gegenbauer-solvable quantum chain model By: Znojil, Miloslav

%\bibitem{NIP}
M. Znojil,
%"Non-Hermitian interaction representation and its
%%use in relativistic quantum mechanics."
%
Ann. Phys. (NY) 385, 162 (2017).
%%pp. 162 - 179 doi: 10.1016/j.aop.2017.08.009 (arXiv:1702.08493v2)

\bibitem{Geyer}
F. G. Scholtz, H. B. Geyer and F. J. W. Hahne, Ann. Phys. (NY) 213,
74 (1992).

\bibitem{Nimrod}
N. Moiseyev, Non-Hermitian Quantum Mechanics (CUP, Cambridge, 2011).


\bibitem{Uweb}
M. Hiller, T. Kottos and A. Ossipov, Phys. Rev. A, 73  063625
(2006).



\bibitem{Uwec}
%Annals of Physics Volume 330, March 2013, Pages 142-159 Annals of
%Physics Scaling behavior and phase diagram of a PT-symmetric
%non-Hermitian Bose–Hubbard system Author links open overlay
%panel
L. Jin and Z. Song, Ann. Phys. (NY) 330, 142 (2013).
% Show more https://doi.org/10.1016/j.aop.2012.11.017



\bibitem{Stone}
M. H. Stone,
%M. H. (1932), "On one-parameter unitary groups in Hilbert Space",
Ann. Math. 33, 643  (1932).
%–648, doi:10.2307/1968538


%
%\bibitem{SIGMA}
%%M. Znojil,
%%%Time-dependent version of cryptohermitian quantum theory
%%Phys. Rev. D 78, 085003  (2008);
%%%(doi: 10.1103/PhysRevD.78.085003)
%%%(arXiv:0809.2874v1 [quant-ph] 17 Sep 2008)
%%
%M. Znojil, SIGMA 5, 001 (2009) (e-print overlay: arXiv:0901.0700).
%%
%%M. Znojil,
%%%"Non-Hermitian interaction representation and its use in
%%%relativistic quantum mechanics."
%%Ann. Phys. (NY) 385, 162 (2017).
%%% pp.
%%%162 - 179 doi: 10.1016/j.aop.2017.08.009 (arXiv:1702.08493v2)
%


%
%



\bibitem{Feshbach}
H. Feshbach,
% H. Unified theory of nuclear reactions.
Ann. Phys. (NY) 5, 357
%–390.
(1958).
%
%(a) Feshbach H. A unified theory of nuclear reactions II. Ann Phys (NY) 1962;19:287–
%313;

\bibitem{Feshbachb}
P.-O. L\"{o}wdin,
% P-O. Studies in perturbation theory. IV. Solution of eigenvalue problem
%by projection operator formalism.
J. Math. Phys. 3, 969
%–982.
(1962).

\bibitem{Ingrid}
I. Rotter,
J. Phys. A: Math. Theor. 42, 153001  (2009).


%
%\bibitem{xx18} M. Philipp, P. v. Brentano, G. Pascovici, A. Richter.
%Frequency and width crossing of two interacting resonances in a
%microwave cavity. Phys Rev E 62:1922–1926, 2000.
%doi:10.1103/PhysRevE.62.1922.
%%
%%\bibitem{xx19} C. Dembowski, H.-D. Gräf, H. L. Harney, et al.
%%Experimental observation of the topological structure of
%%exceptional points. Phys Rev Lett 86:787–790, 2001.
%%doi:10.1103/PhysRevLett.86.787.
%%
%%\bibitem{xx20} C. Dembowski, B. Dietz, H.-D. Gräf, et al.
%%Observation of a chiral state in a microwave cavity.
%%Phys Rev Lett 90:034101, 2003.
%%doi:10.1103/PhysRevLett.90.034101.


\bibitem{xx22}
T. Stehmann, W. D. Heiss and F. G. Scholtz,
% Observation
%of exceptional points in electronic circuits.
J. Phys. A: Math. Gen. 37, 7813 (2004).
%2004. doi:10.1088/0305-4470/37/31/012.
%
%\bibitem{xx21} B. Dietz, T. Friedrich, J. Metz, et al. Rabi oscillations
%at exceptional points in microwave billiards. Phys Rev
%E 75:027201, 2007. doi:10.1103/PhysRevE.75.027201.
%
%
%\bibitem{xx23} M. Lawrence, N. Xu, X. Zhang, et al. Manifestation
%of PT symmetry breaking in polarization space with
%terahertz metasurfaces. Phys Rev Lett 113:093901,
%2014. doi:10.1103/PhysRevLett.113.093901.

\bibitem{xx24}
T. Gao, E. Estrecho, K. Y. Bliokh et al,
%. Observation
%of non-Hermitian degeneracies in a chaotic exciton-polariton
%billiard.
Nature 526, 554
%–558,
(2015).
%. doi:10.1038/nature15522.

%\bibitem{xx25} J. Doppler, A. A. Mailybaev, J. Böhm, et al.
%Dynamically encircling an exceptional point for
%asymmetric mode switching. Nature 537(7618):76–79,
%2016. doi:10.1038/nature18605.

\bibitem{xx26}
H. Xu, D. Mason, L. Jiang and J. G. E. Harris,
%Topological energy transfer in an optomechanical system with
%exceptional points.
Nature 537, 80
%–83,
(2016).
%doi:10.1038/nature18604.



\bibitem{Cart2}
R. El-Ganainy, K. G. Makris, D. N. Christodoulides
 and Z. H.
Musslimani,
%Z. H. Theory of coupled optical PT-symmetric structures.
Opt. Lett. 32, 2632
%- 2634
(2007);

Z. Lin, A. Pick, M. Lon\v{c}ar and A. W.
Rodriguez, Phys. Rev. Lett. 117, 107402
%Enhanced Spontaneous Emission at Third-Order Dirac Exceptional
%Points in Inverse-Designed Photonic Crystals By:Lin, Z (Lin, Zin)[ 1
%] ; Pick, A (Pick, Adi)[ 2,3 ] ; Loncar, M (Loncar, Marko)[ 1 ] ;
%Rodriguez, AW (Rodriguez, Alejandro W.)[ 4 ] PHYSICAL REVIEW LETTERS
%Volume: 117  Issue: 10 Article Number: 107402 DOI:
%10.1103/PhysRevLett.117.107402 Published:AUG 30
(2016);

%Enhanced sensitivity at higher-order exceptional points By:Hodaei, H
%(Hodaei, Hossein)[ 1 ] ; Hassan, AU (Hassan, Absar U.)[ 1 ] ;
%Wittek, S (Wittek, Steffen)[ 1 ] ; Garcia-Gracia, H (Garcia-Gracia,
%Hipolito)[ 1 ] ; El-Ganainy, R (El-Ganainy, Ramy)[ 2,3 ] ;
%Christodoulides, DN (Christodoulides, Demetrios N.)[ 1 ] ;
%Khajavikhan, M (Khajavikhan, Mercedeh)[ 1 ] NATURE Volume: 548
%Issue: 7666  Pages: 187-+ DOI: 10.1038/nature23280 Published:AUG 10
%2017
%
%Enhanced sensitivity at higher-order exceptional points
H. Hodaei, A. U.
Hassan, S. Wittek, H. Garcia-Gracia, R. El-Ganainy,
D. N. Christodoulides and M. Khajavikhan,
% Nature, 2017 - nature.com …
Nature 548, 187
%- 191
(2017);

J. Schnabel, H. Cartarius, J. Main, G. Wunner and W. D. Heiss,
% PT -symmetric waveguide
%system with evidence of a third-order exceptional point.
Phys. Rev. A 95,
053868 (2017).
%doi:10.1103/PhysRevA.95.053868.


\bibitem{Cart}
W. D. Heiss and G. Wunner,
%. A model of three coupled
%wave guides and third order exceptional points. J Phys
%A 49(49):495303, 2016.
%doi:10.1088/1751-8113/49/49/495303
 J. Phys. A: Math. Theor. 49, 495303 (2016);

J. Schnabel, H. Cartarius, J. Main,
G. Wunner and W. D.
Heiss,
Acta Polytech. 57, 454 (2017).
%–461, 2017
%
%http://ojs.cvut.cz/ojs/index.php/ap SIMPLE MODELS OF THREE COUPLED
%PT -SYMMETRIC WAVE GUIDES ALLOWING FOR THIRD-ORDER EXCEPTIONAL
%POINTS
%with a PT -symmetric
%distribution of gain and loss


\end{thebibliography}
\end{document}